\documentclass[aps,prl,twocolumn,superscriptaddress,showpacs]{revtex4}

\usepackage{amsmath}
\usepackage{amssymb}
\usepackage{graphicx}
\usepackage{dcolumn}
\usepackage{bm}
\usepackage{psfrag}
\usepackage{units}
\usepackage{slashed}
\usepackage{bm}

\begin{document}

\title{Test of the Atiyah-Singer Index Theorem for Fullerene with a Superconducting Microwave Resonator}

\author{B.~Dietz}
\email{dietz@ikp.tu-darmstadt.de}
\affiliation{Institut f{\"u}r Kernphysik, Technische Universit{\"a}t
Darmstadt, D-64289 Darmstadt, Germany}

\author{T.~Klaus}
\affiliation{Institut f{\"u}r Kernphysik, Technische Universit{\"a}t
Darmstadt, D-64289 Darmstadt, Germany}

\author{M.~Miski-Oglu}
\affiliation{Institut f{\"u}r Kernphysik, Technische Universit{\"a}t
Darmstadt, D-64289 Darmstadt, Germany}

\author{A.~Richter}
\affiliation{Institut f{\"u}r
Kernphysik, Technische Universit{\"a}t Darmstadt, D-64289 Darmstadt,
Germany}

\author{M. Bischoff}
\affiliation{Theoriezentrum, Institut f{\"u}r Kernphysik, Technische Universit{\"a}t
Darmstadt, D-64289 Darmstadt, Germany}

\author{L. von~Smekal}
\affiliation{Theoriezentrum, Institut f{\"u}r Kernphysik, Technische Universit{\"a}t
Darmstadt, D-64289 Darmstadt, Germany}
\affiliation{Institut f{\"u}r Theoretische Physik, Justus-Liebig-Universit{\"a}t
Gie\ss en, D-35392 Gie\ss en, Germany}

\author{J.~Wambach}
\affiliation{Theoriezentrum, Institut f{\"u}r Kernphysik, Technische Universit{\"a}t
Darmstadt, D-64289 Darmstadt, Germany}

\date{\today}

\begin{abstract}

Experiments have been performed using a spherical superconducting microwave resonator that simulates the geometric structure of the C$_{60}$ fullerene molecule. The objective was to study with very high resolution the exceptional spectral properties emerging from the symmetries of the icosahedral structure of the carbon lattice. In particular, the number of zero modes has been determined to test the predictions of the Atiyah-Singer index theorem, which relates it to the topology of the curved carbon lattice. This is, to the best of our knowledge, the first experimental verification of the index theorem.

\end{abstract}

\maketitle
{\it Introduction.}--- The spectrum of graphene, a monolayer of carbon (C) atoms arranged on a hexagonal lattice, has been the focus of extensive theoretical~\cite{Beenakker2008,Castro2009} and experimental studies~\cite{Ponomarenko2008}. Its universal properties were often also investigated experimentally in analog systems, so-called 'artificial graphene'~\cite{Polini2013a}, e.g., in our group in photonic crystals~\cite{Bittner2010,Kuhl2010,Bellec2013,Rechtsman2013,Khanikaev2013,Dietz2015}. Moreover, theoretically much attention has been devoted to curved graphene structures like fullerene molecules~\cite{Kroto1985,Manousakis1991,Gonzalez1992,Gonzalez1993,Kolesnikov2006,Vozmediano2010} and the connection between their spatial symmetries and electronic properties. Here, the most famous example is the C$_{60}$ molecule. It consists of 60 carbon atoms at the vertices of a truncated icosahedron and has the shape of a soccer ball. Concerning the spectral properties of fullerenes the number of near-zero modes, i.e., of electronic states with excitation energies close to zero, have been of particular interest since they determine the electrical conductivity. In~\cite{Pachos2007,Pachos2007a} an index theorem has been derived that allows the computation of the number of such near-zero modes from the topology of the surface. It was deduced from the renowned Atiyah-Singer index theorem~\cite{Atiyah1963,Atiyah1968,Atiyah1968a,Atiyah1968b} which states that the analytic index of an elliptic differential operator on a compact manifold equals the topological one, in other words, that there is a connection between the number of zero modes of the operator and the topology of the manifold on which it is defined. The aim of the high-resolution experiments presented in this letter was to test these predictions in experiments with a superconducting microwave resonator of the same topology as the C$_{60}$ molecule.      

First we briefly review the salient features of graphene and fullerenes  and outline the derivation of the index theorem from~\cite{Pachos2007,Pachos2007a} for deformed graphene sheets. We then describe the experimental setup and compare the results of the measurements to the predictions from the index theorem and to tight-binding model (TBM) calculations. These allow us to study the approach to the thermodynamic limit of an infinite number of carbon atoms. 

{\it Graphene, fullerenes and the Atiyah-Singer theorem.}--- 
The honeycomb structure of graphene is formed by two interpenetrating triangular sublattices. As a consequence, at half filling the Fermi surface in graphene reduces to two independent points in the first Brillouin zone, the so-called 'Dirac points', denoted by ${\bf K_+}$ and ${\bf K_-}$~\cite{Beenakker2008,Castro2009} that are conical intersections of the valence and the conduction band. Low energy excitations within the cone regions around ${\bf K_\pm}$ have a linear dispersion with a slope given by the Fermi velocity $v_F$. On an infinte graphene sheet they are therefore described by a Dirac Hamiltonian for massless spin-1/2 quasiparticles consisting of partner Hamiltonians  
\begin{equation}
H_\pm=\pm v_F \, \sigma^\alpha q_\alpha
\end{equation}
which describe excitations with momentum ${\bf q}=(q_x,q_y)$ in each of the two Dirac cones around ${\bf K_\pm}$. The Pauli matrices $\sigma^\alpha $ with $\alpha = x,y$ act on the two sublattice components of the excitations, combined in two-dimensional spinors and hence referred to as quasi-spin. Both cones together then yield a four-component Dirac equation~\cite{Beenakker2008}. 

Fullerene molecules can be constructed by introducing positive curvature into an initially flat graphene sheet~\cite{Lammert2004}. The bending is realized by replacing hexagons by pentagons, ensuring at the same time that the lattice is not stretched and each C atom keeps three neighbors. To determine the number of pentagons $n_5$ necessary to generate a spherical fullerene molecule with $n_6$ hexagons one uses the Euler formula~\cite{Millman1977} which relates the number of vertices $V$, of edges $E$, faces $F$ and open ends $N_\mathrm{open}$ of an arbitrary two-dimensional lattice to the genus $g$ of the surface formed by it, via the Euler characteristic  
\[ \chi = V-E+F = 2 ( 1-g) - N_\mathrm{open}  \, . \]
For a lattice of pentagons and hexagons, $V=(5n_5+6n_6)/3$, $E=(5n_5+6n_6)/2$ and $F=n_5+n_6$, this gives $\chi =n_5/6$. Without open ends $\chi$ must be an even integer on a closed orientable surface due to the Gau\ss-Bonnet theorem~\cite{Millman1977}. Hence, for a flat graphene sheet with periodic boundary conditions one has $g =1$ for the torus and $n_5=0$, while a sphere with $g=0$ needs $n_5 = 12$ pentagons to avoid open ends. Consequently, fullerenes are grown from the C$_{60}$ molecule by increasing the number of hexagons, i.e., always have twelve pentagons at the same relative positions. This also applies to the thermodynamic limit, and one expects that their low-energy electronic excitations are described by a Dirac equation on a sphere. 

\begin{figure}[!ht]
\includegraphics[width=\linewidth]{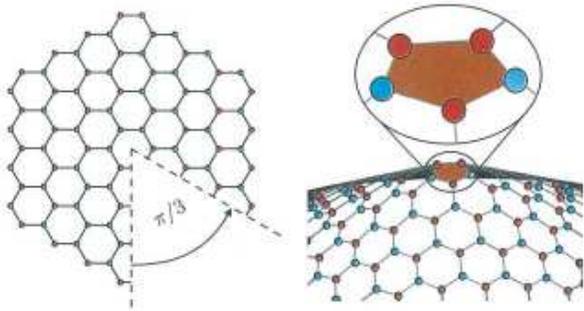}
\caption{(Color online) Left panel: A two-dimensional graphene sheet. The red and the blue dots mark the two independent triangular sublattices. In order to form a curved sheet which contains one pentagon, a $\pi/3$ segment is cut out from the sheet. Right panel: The conically deformed graphene sheet with the pentagon at the apex.} 
\label{fig1}
\end{figure}

To introduce a pentagon into the honycomb lattice, a $\pi /3$ sector is cut out and then the edges are glued together~\cite{Gonzalez1992,Gonzalez1993,Pachos2007,Pachos2007a} as illustrated in Fig.~\ref{fig1}.
Thereby a pentagon is created at the apex of the emerging cone. Along the seam, two C atoms from the same triangular sublattice, e.g., the red ones in Fig.~\ref{fig1}, are connected. This results in a coupling of the Dirac operators associated with the $\bf K_\pm$ points. Indeed, when the four-dimensional spinor associated with the Dirac equation of the flat graphene sheet is transported around the apex by an angle $2\pi$, it is forced to jump at the seam from a red site to another red one instead of to a blue one. It thus acquires a non-trivial phase, which can be accounted for by introducing a non-Abelian gauge field $A_\mu$ in the Hamiltonian which yields a flux of $ (\pi/2) \, \tau^y $
when integrated along a closed loop around the apex. The Pauli matrix $\tau^y$ thereby couples the ${\bf K_+}$ and ${\bf K_-}$ spinor components~\cite{Beenakker2008}. This description entails the existence of a ficitious magnetic monopole inside the surface. In the case of fullerenes it is located at the center of the spherical molecule, yielding a flux of $1/8$ through each of the twelve pentagons. Thus the total magnetic monopole charge inside the sphere equals 3/2. In addition to that, analogous to the daily rotation of Foucault's pendulum, a deficit angle of $\pi/3$ arises when moving a frame along a loop around the apex. It is here described by a quasi-spin connection $Q_\mu$ with circulation $ -({\pi}/{6})\,  \sigma^z $ around the apex.  

The coupling of the ${\bf K_\pm}$ spinor components in the resulting four-dimensional Dirac equation can be removed by a rotation, which leads to two independent two-dimensional Dirac equations denoted by $l=1,2$ \cite{Pachos2007a},
\begin{equation}
\slashed D^l\psi^l= v_F \, \sigma^\alpha e_\alpha^\mu\left(q_\mu -iQ_\mu-iA_\mu^l\right)\psi^l= E\psi^l\, ,
\label{DE}
\end{equation}    
where $e_\alpha^\mu$ is the Zweibein in the tangent plane of the surface, and 
$A^l_\mu$ are the components of $A_\mu$  in the rotated basis with circulation  
$\pm \pi/2 $ for $l= 1$ and $2$, respectively, where $\vec A^l$ is now an Abelian gauge field.

The four-dimensional Dirac equation obtained from Eq.~(\ref{DE}) provides a good description of the low-energy excitations of the C molecules~\cite{Gonzalez1992,Gonzalez1993,Kolesnikov2006}. It yields the long-wavelength excitations of the deformed graphene sheet in the vicinity of the Dirac points, and thus also the zero modes that we are interested in. For the fullerenes the Dirac operators $\slashed D^l$ are elliptic and defined on a compact surface. Hence the Atiyah-Singer index theorem~\cite{Atiyah1963,Atiyah1968,Atiyah1968a,Atiyah1968b} applies. Ten years after its first formulation a new proof was provided based on the heat equation~\cite{Atiyah1973} which was later employed  for the derivation of an index theorem for graphene sheets deformed by pentagons and heptagons \cite{Pachos2007,Pachos2007a} as briefly reviewed in the following. 

Each Dirac operator in Eq.~(\ref{DE}) can be written in terms of off-diagonal 
partner operators $P$ and $ P^\dagger$ \cite{Stone1984}, so $(\slashed D^l)^2$ contains only diagonal operators $PP^\dagger$ and $P^\dagger P$ that have the same number of zero modes as $P^\dagger$ and $P$, respectively. Furthermore, the non-zero eigenvalues of $PP^\dagger$ and $P^\dagger P$ are identical. The analytic index of $\slashed D^l$ is given by the difference of the numbers of zero modes of $P$ and $P^\dagger$ denoted by $\nu_\pm$, respectively, i.e.,  $\mathrm{index}(\slashed D^l)=\nu_+-\nu_-$ \cite{Pachos2007,Pachos2007a}.
More importantly, however, this index is related to the total flux of the effective gauge field via the Atiyah-Singer index theorem \cite{Pachos2007}, 
\begin{equation}
\mathrm{index}(\slashed D^l)=\frac{1}{2\pi}\iint_\Omega {\mathcal{F}}^l {\rm d}\Omega\, .
\label{AS}
\end{equation}
The integral is taken over the compact surface $\Omega$ and ${\mathcal{F}}^l=\partial\wedge A^l$ are the field strengths associated with the now Abelian gauge potentials $A^l$. Stokes' theorem then implies from the closed loops around each apex that
\begin{equation}
\frac{1}{2\pi}\iint_{\Omega}{\mathcal{F}}^l {\rm d}\Omega =\frac{1}{2\pi}n_5\oint A^l_\mu \, {\rm d}s^\mu \equiv\pm\frac{3}{2}\chi\, .
\end{equation}
The Euler formula thus leads to the Atiyah-Singer index theorem for fullerenes in the form, $\mathrm{index}(\slashed D^l)=\pm 3(1-g)$. In two dimensions either $\nu_+$ or $\nu_-$ vanish. Hence, the index theorem provides the number of zero modes~\cite{Pachos2007,Pachos2007a}. 

The total number of zero modes is the sum of those of the subsystems corresponding to $l=1,2$. Consequently, according to the index theorem, the zero modes of the four-dimensional Dirac operator for spherical fullerenes correspond to two triplets. The same result has been obtained in a continuum model for the low-energy electronic states of icosahedral fullerenes~\cite{Gonzalez1992,Gonzalez1993,Kolesnikov2006,Vozmediano2010}. We emphasize, however, that the eigenvalues of the near-zero modes tend to zero, i.e., coincide with the energy at the Dirac points, only in the thermodynamic limit of an infinity number of $C$ atoms. In a sufficiently large but finite fullerene molecule they are expected to lie much closer to the Dirac energy than all the other ones.
  
{\it Experimental setup and resonance spectra.}---
Hitherto, experiments have been performed with flat, superconducting microwave resonators, so-called 'microwave billiards'~\cite{Richter1999} to address problems from the fields of quantum chaos~\cite{Haake2001,StoeckmannBuch2000} and compound nucleus reactions~\cite{Mitchell2010}. In this context, the equivalence of the Helmholtz equation and the non-relativistic Schr\"odinger equation of the corresponding quantum billiard is exploited which holds below a maximum microwave frequency $f_{\rm max}=c/(2d)$ with $c$ the velocity of light and $d$ the height of the billiard. Consequently, the eigenvalues of a quantum billiard can be obtained experimentally from the eigenfreqencies of the microwave billiard of corresponding shape. Recently, we realized experiments with superconducting microwave Dirac billiards and studied universal spectral properties of graphene sheets~\cite{Wurm2011} with unprecedented accuracy~\cite{Bittner2012,Dietz2015,Dietz2015a}. 

\begin{figure}[!ht]
\includegraphics[width=0.6\linewidth]{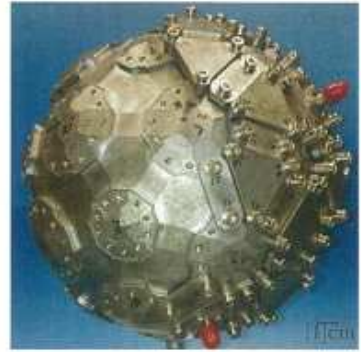}
\caption{(Color online) Lead plated fullerene billiard used in the experiments. In the left part the small plates that cover the circular cavities and the rectangular channels were removed. The red caps protect the antenna ports. The billiard is superconducting below T$_c$=7.2~K.}
\label{fig2}
\end{figure}

The aim of the experiments presented here was the investigation of the universal spectral properties of the fullerene C$_{60}$ molecule attributed to its lattice structure and to determine the number of zero modes which, according to the Atiyah Singer index theorem solely depends on the number of pentagons. For this we use a system exhibiting the same topological properties, namely a quantum fullerene billiard on a sphere, consisting of a network of 60 circular billiards at the positions of the C atoms connected by three of the altogether 90 straight leads with three adjacent ones. We studied them experimentally by using instead of a planar, superconducting microwave (Dirac) billiard a cavity, which is imprinted on a sphere. The microwave fullerene billiard displayed in Fig.~\ref{fig2} was constructed by milling a total of 60 circular cavities (vertices) and 90 rectangular channels (edges) out of a brass sphere and then closing them with small triangular brass plates of 5~mm thickness and 3~mm thick rectangular ones, respectively. Before the parts were screwed together, they were covered with lead, which is superconducting below $T_c=7.2$~K.
The diameter of the sphere of 160~mm was limited by the size of the liquid Helium cryostat in which the resonator was cooled down to 4.2~K in order to attain superconductivity. The radius of the circular cavities was 12~mm, the widths of the waveguides 14~mm, before lead coating them. Thus the cutoff frequency for the first propagating mode in the latter is $f_c^1\gtrsim 10.714$~GHz. In total, 8 antennas were attached to the triangular plates. Two, covered with red caps, are visible in Fig.~\ref{fig1}. The height of the resonator was 3~mm corresponding to $f_{\rm max}=50$~GHz.   

\begin{figure}[!ht]
\includegraphics[width=\linewidth]{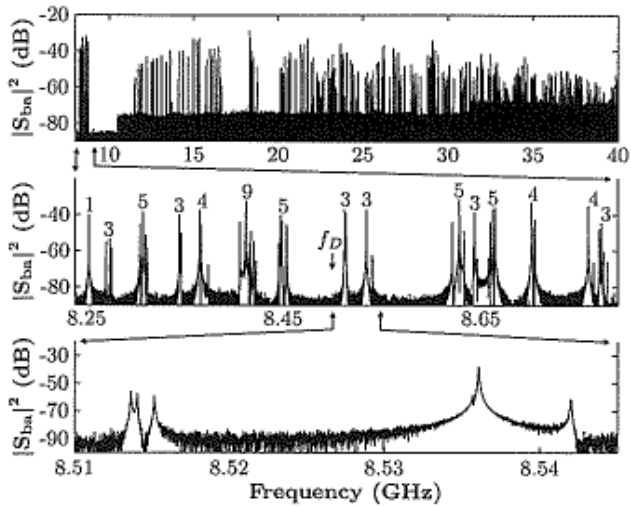}
\caption{Transmission spectrum of the fullerene billiard (upper panel) up to 40 GHz. The first band ranges from 8.254 to 8.779~GHz. It contains $60$ resonances, that are separated into 15 groups with the number of resonances indicated in the middle panel, which shows a zoom into it. The Dirac frequency $f_D=8.504$~GHz is marked by an arrow. The two triplets of interest (the zero modes) are clearly resolved and shown in the lower panel.}
\label{fig3}
\end{figure}

For the measurement of the transmission spectrum shown in the upper panel of Fig.~\ref{fig3} microwave power was coupled into the microwave billiard via antenna a and the output signal was received at antenna b, with a and b denoting two of the 8 antennas. A vectorial network analyzer determined the relative phase and amplitude of the output and input signals, thus yielding the scattering matrix element $S_{\rm ba}$. The smallest resonance frequency equals $f=8.254$~GHz, so we show the spectrum from 8 - 40~GHz. Due to the high-quality factor $Q>10^5$ of the resonator, all resonances could be resolved in that frequency range. We concentrate our discussion here on the region between 8.254 and 18.801~GHz. 
The spectrum exhibits three distinct bands, containing 60, 210 and 90 resonances, respectively, in the frequency intervals [8.254,\, 8.779]~GHz, [11.492,\, 16.657]~GHz and [18.312,\, 18.801]~GHz. They are located around the eigenfrequencies $\tilde f_1\simeq 8.4$~GHz, $\tilde f_2\simeq 13.5$~GHz and $\tilde f_3\simeq 18.5$~GHz of the first three quasibound states in an open circular billiard of the same size as the cavities in the resonator with openings at the positions of the waveguides. The modes excited inside the circular cavities resemble within a given band the corresponding mode in the open circular billiard, and are described by $J_0$, $J_1$ and $J_2$ Bessel functions in the first, second and third band, respectively. In the latter two cases they are twofold degenerate due to the mirror symmetry. Note, that the circular cavities exhibit no threefold symmetry, because each of them is part of one pentagon and two hexagons and the internal angles differ. 

The first band is located well below the cutoff frequency of the waveguide. Consequently, the electric field modes excited inside the circular cavities are only weakly coupled to those in the neighboring ones. The resonance frequencies in the second band are above the cutoff frequency. Accordingly, the modes in the cavities are coupled via the modes inside the waveguides, and thus mimick a situation where the C atoms are coupled to the neighboring ones via an extra atom, thus explaining the number of resonances in this band. The third band is still below the frequency $f_c^2\gtrsim 20.143$~GHz of the second propagating mode in the waveguides. As a result, the number of possible mode configurations is restricted due to the symmetry properties of the modes excited inside the cavities, that prefereably couple to the second excited mode inside the waveguide. Above 20.232~GHz, i.e., beyond $f_c^2$, several bands are intertwined.
 
In summary, only the first band can be used to model the situation in the fullerene C$_{60}$ molecule. The middle panel of Fig.~\ref{fig3} shows a magnification of it. Fifteen groups of nearly degenerate resonances are clearly visible. The number of resonances identified in each of them is indicated and coincides with the degrees of degeneracy predicted on the basis of group theoretical considerations for the eigenfrequencies because of its truncated icosahedral structure~\cite{Manousakis1991,Dresselhaus1996}. In the group with degeneracy degree 9, in fact, the energy values of 5 and 4 degenerate eigenfrequencies, respectively, are accidentally the same. We emphasize, that we were only able to identify \emph{all} 60 resonance frequencies because the degeneracies were lifted. The reason is that the symmetries of the resonator structure were slightly perturbed due to the presence of the antennas and unavoidable marginal inhomogeneities in the lead coating. The influence of the former turned out to be negligible for sufficiently short antennas. The effect of the latter on the size of the splittings of the nearly degenerate resonance frequencies was tested by smoothing the surface of the resonator which indeed induced a reduction of the splittings in each group. The pair of triplets visible in the middle panel of Fig.~\ref{fig3} and shown in a further magnification of the spectrum in the lowest panel, is closest to the Dirac frequency at $f_D=8.504$~GHz which was determined as described below. The zoom demonstrates the high resolution necessary to resolve the 6 resonances. These are the 6 modes conjectured by the Atiyah-Singer index theorem that we were looking for. As is discussed next they are corroborated also by TBM calculations.       

\begin{figure}[!ht]
\includegraphics[width=\linewidth]{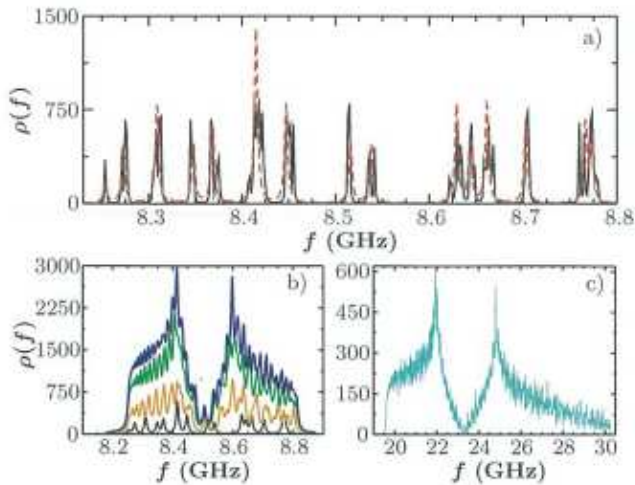}
\caption{(Color online) Panel a): Resonance density determined from the resonance frequencies of the first band (middle panel in Fig.~\ref{fig3}) of the fullerene billiard (black solid line) compared to the TBM calculations (red dashed line). 
Panel b): Resonance densities computed within the TBM for C$_{60}$, C$_{240}$, C$_{540}$ and C$_{720}$ (lowermost to upmost curve). Panel c): Experimentally determined resonance density of a rectangular Dirac billiard~\cite{Dietz2013,Dietz2015}.}
\label{fig4}
\end{figure}

{\it Tight-binding model description of the spectra.}---
The eigenvalues of the C$_{60}$ molecule have been computed previously using the TBM~\cite{Manousakis1991}, however, a stringent test of its applicability was missing. Given the experimental results on the eigenfrequencies in the first band of the fullerene resonator we are now in a position to check the validity of the TBM in detail. As stated above, the modes excited in the 60 cavities are weakly coupled, which is an essential prerequisit for the applicability of the TBM. Detailed calculations showed a quantitative agreement between the computed and the measured frequencies only when including next-nearest, and second and third-nearest neighbor couplings with strengths $t_1,\, t_2$ and $t_3$, respectively. This yielded for the frequency of the isolated cavities $f_0=8.515$~GHz and the coupling parameters $t_1=-0.0929$~GHz, $t_2=0.0035$~GHz and $t_3=0.0005$~GHz. The eigenvalues deduced from the TBM appear as 15 groups of degenerate eigenvalues with the same multiplicities as the resonances in the spectrum shown in the middle panel of Fig.~\ref{fig3}. An even better agreement was achieved by taking into account the fact that, due to the inhomogeneities in the lead coating, the radii of the cavities are slightly different. In order to estimate the deviations thus induced in $f_0$, we used the fact, that  $f_0$ is given by the first zero of the $J_0$ Bessel function, $J_0(kR)=0$, for a circular cavity of radius $R$. We inserted for each of the 60 cavities the measured radius and replaced $f_0$ by the individual values. Thereby, the degeneracies were removed. In panel a) of Fig.~\ref{fig4} we compare the resonance density $\rho (f)=\sum_i\delta(f-f_i)$ obtained for the frequencies $f_i$ in the first band (black full line) with the TBM result (red dashed line). For display purposes we have replaced the $\delta$ functions by Lorentzians of finite width of $\Gamma=2$~MHz. The good agreement reassures the applicability of the TBM and thus justifies its use for further numerical studies with larger fullerene molecules with the parameters determined from the experiment.  

Panel b) of Fig.~\ref{fig4} shows a comparison between the resonance densities of the C$_{60}$, C$_{240}$, C$_{540}$ and C$_{720}$ molecules in ascending order. Here, we used Lorentzians of width $\Gamma=5$~MHz for all cases. As stated above, all molecules contain the same number of pentagons whereas the number of hexagons increases. The resonance densities should thus resemble more and more that of a graphene sheet. They exhibit a minimum bounded by two increasingly sharp peaks, that evolve into van Hove singularities in the limit of an infinite number of atoms~\cite{VanHove1953,Dietz2013}. A comparison of the resonance densities with that of a rectangular graphene sheet with periodic boundary conditions (panel c) of Fig.~\ref{fig4}), which has been obtained in measurements with a microwave Dirac billiard~\cite{Dietz2015}, shows that they resemble for large fullerenes. In contradistinction to the latter, however, the resonance densities of the fullerenes all exhibit a peak of similar size located at the minimum which is due to the two triplets of zero modes. This remains true in the limit of an infinite number of atoms~\cite{Gonzalez1992,Gonzalez1993,Kolesnikov2006,Vozmediano2010} and is thus a distinct feature of the spatial curvature and topology. According to the Atiyah-Singer index theorem a plane graphene sheet with periodic boundary conditions should not exhibit any zero modes. This is observed in panel c). Zero modes are expected only, if part of the graphene sheet is terminated with zigzag edges~\cite{Wurm2011}. The associated states are called edge states, because their wave functions vanish everywhere except at these edges~\cite{Kuhl2010,Bittner2012,Dietz2015}. We have also computed the wave functions of the fullerene molecules under consideration using the TBM and found, that those of the 6 zero modes are localized at the pentagons, which may be considered to be equivalent to zigzag edges within the hexagon network. 

\begin{figure}[!ht]
\includegraphics[width=\linewidth]{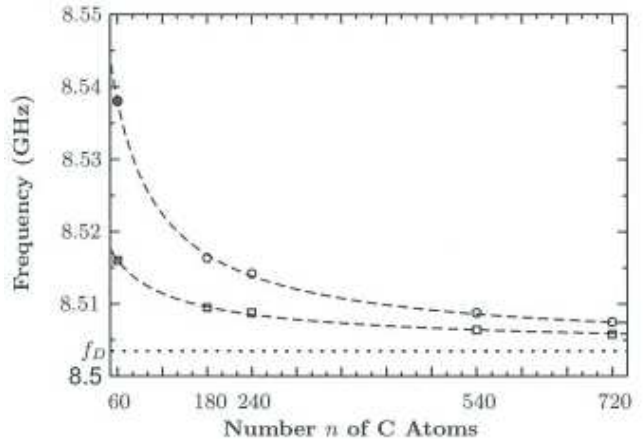}
\caption{(Color online) Calculated frequencies of the zero modes of the fullerenes C$_{60}$, C$_{180}$, C$_{240}$, C$_{540}$ and C$_{720}$ vs. the number $n$ of C atoms. All pairs of triplets (squares and circles) are located slightly above the Dirac frequency (dotted line). The experimental values are marked by a red filled circle and square. The dashed lines correspond to fits of the function $f(n)=f_D+a/n^b$ to the data points yielding for the Dirac frequency $f_D=8.504$~GHz, $a=1.188$~GHz, $b=0.661$ for the lower zero modes and $a=1.211$~GHz, $b=0.869$ for the upper ones.} 
\label{fig5}
\end{figure}

The central frequencies of the two triplets, marked by squares for the one close to the Dirac frequency (dotted line) and by a circle for the one further away are displayed in Fig.~\ref{fig5} as a function of the number $n$ of C atoms. As is clearly visible, the distance between the triplets decreases with increasing size of the fullerene molecule and both approach the Dirac frequency. This behavior is well fitted by a function $f(n)= f_D+a/n^b$ yielding  the parameter values given in the caption of Fig.~\ref{fig5}, and for the Dirac frequency finally a value of $f_D=8.504$~GHz. Note that this is essentially the only way to determine the Dirac frequency of a C$_{60}$ molecule.

{\it Conclusions.}--- The lowest 60 eigenvalues of a C$_{60}$ fullerene were determined in high-precision experiments using a superconducting microwave billiard of corresponding shape. They appear in 15 groups of nearly degenerate ones, where the multiplicity coincides with that determined based on the group theory of the truncated icosahedral structure of C$_{60}$. We have demonstrated in TBM calculations for spherical fullerene molecules of increasing size that the two triplets of resonances detected close to the Dirac frequency correspond to the triplets of zero modes predicted by the Atiyah-Singer index theorem, and thus provided to the best of our knowledge the first experimental test of it. The exact value of the Dirac frequency was obtained as the asymptotic value attained by the frequencies of the triplets in the limit of an infinite number of atoms. 

This work was supported by the DFG within the Collaborative Research Center 634.

\end{document}